\documentclass[sigconf,nonacm]{acmart}
\usepackage{subfig}
\usepackage{hyperref}
\usepackage{multirow}
\usepackage{balance}

\AtBeginDocument{%
  }

\begin{document}

\title{Multi-modal Atmospheric Sensing to Augment Wearable IMU-Based Hand Washing Detection}

\author{Robin Burchard}
\email{robin.burchard@uni-siegen.de}
\orcid{0000-0002-4130-5287}
\affiliation{%
  \institution{University of Siegen}
  \streetaddress{Hölderlinstr. 3A}
  \city{Siegen}
  \country{Germany}
  \postcode{57076}
}
\author{Kristof Van Laerhoven}
\email{kvl@eti.uni-siegen.de}
\orcid{0000-0001-5296-5347}

\affiliation{%
  \institution{University of Siegen}
  \streetaddress{Hölderlinstr. 3A}
  \city{Siegen}
  \country{Germany}
  \postcode{57076}
}

\begin{abstract}
Hand washing is a crucial part of personal hygiene. Hand washing detection is a relevant topic for wearable sensing with applications in the medical and professional fields. Hand washing detection can be used to aid workers in complying with hygiene rules. Hand washing detection using body-worn IMU-based sensor systems has been shown to be a feasible approach, although, for some reported results, the specificity of the detection was low, leading to a high rate of false positives. In this work, we present a novel, open-source prototype device that additionally includes a humidity, temperature, and barometric sensor. We contribute a benchmark dataset of 10 participants and 43 hand-washing events and perform an evaluation of the sensors' benefits. Added to that, we outline the usefulness of the additional sensor in both the annotation pipeline and the machine learning models. By visual inspection, we show that especially the humidity sensor registers a strong increase in the relative humidity during a hand-washing activity. A machine learning analysis of our data shows that distinct features benefiting from such relative humidity patterns remain to be identified.
\end{abstract}



\maketitle

\section{Introduction}

Hand washing is very relevant for our personal health because it effectively reduces the amount of bacteria, viruses, and other pathogens that we spread \cite{burton_effect_2011}.
Added to protecting our own health, effective hand washing can also reduce the spread of illnesses to other human beings or animals. Many professional applications require food-safe or sterile environments, for which hand washing and e.g. alcohol-based hand sanitizers can be used. Automatically measuring the frequency, duration, and quality of hand washing therefore could be beneficial in a multitude of applications.
There are also prospective applications in the field of mental health. Hand washing detection has for instance been proposed as a measure to aid in the treatment of people with Obsessive Compulsive Disorder (OCD) with washing compulsions \cite{wahl_automatic_2022, wahl_real-time_2023, burchard_washspot_2022}.

Spotting hand washing in the real world and distinguishing it from everyday activities and activities of daily living (ADL) has been shown to be difficult and inaccurate using only data from wrist-worn IMU devices. However, additional cues from the environment such as Bluetooth beacons near sinks can be used to support the hand washing detection \cite{mondol_harmony_2015, cao_leveraging_2023}.
The current approaches for hand washing detection employ a multitude of sensors, although the most frequently used sensors are RGB(D)-cameras and inertial measurement units (IMUs). One category of sensors that has recently been incorporated in wearable devices remains under-explored, namely atmospheric sensors which allow the capturing of humidity, temperature, and air pressure.

\subsection*{Goals and Contributions}
The goal of this work is to explore the use of additional sensing capabilities and their effect on the detection performance of hand washing detection using wearable devices.
The contributions of our work are the following:
\begin{enumerate}
    \item Description of the development and evaluation of an open-source, cost-effective sensor recording device 
    \item Evaluation and comparison of the impact of the added sensors for hand washing detection
    \item Providing an expert-annotated and easy-to-extend dataset and code to reproduce our results and further develop the device and hand washing detection methods.
\end{enumerate}

\section{Related Work}
\label{sec:rel_work}
In the following section, we discuss existing research work on hand washing detection and on the usage of atmospheric sensors in Human Activity Recognition (HAR) in particular. 
Although several recent smartwatches do contain cost-effective miniature sensors such as the Bosch BME280 that sense the humidity of the wearer's surroundings, the modality of humidity is not very prevalent in wearable studies. To the best of the authors' knowledge, there exists no previously published research that combines the use of wearable atmospheric sensors with the goal of hand washing detection. In this work, we focus only on body-worn sensors, as externally placed sensors (e.g. cameras) have several disadvantages for hand washing detection, such as the need to deploy them in sensitive environments such as users' bathrooms, and would need to cover all possible places where users could possibly wash their hands.

\subsection*{Hand washing detection}
For hand washing detection, the most used sensors are inertial measurement unit (IMU) sensors, which contain inertial 3D sensors such as accelerometers and gyroscopes. 
While studies have shown that these sensors on their own can deliver adequate data to detect lab-recorded hand washing and hand washing steps \cite{ivanovs_automated_2020, wang_accurate_2020, wang_you_2021, li_wristwash_2018, lattanzi_unstructured_2022}, no large-scale in-the-wild study exists thus far that can detect hand washing from IMU data alone with close-to-perfect precision and recall \cite{burchard_washspot_2022}. In-lab studies offer higher performance \cite{zhang_detecting_2021}.

Hence, additional sensors or cues could offer a way to make the detection of hand washing more reliable. As an example of popular technology, the Apple Watch can detect hand washing by allegedly using a combination of IMU sensing and the microphone with a proprietary algorithm \cite{hayes_apple_2020, apple_inc_set_2024}. Zhuang et al. also employ an acoustic model \cite{zhuang_detecting_2023} for hand washing detection. Added to that, some studies have used Bluetooth beacons as part of their detection framework \cite{mondol_harmony_2015, cao_leveraging_2023}. However, this only makes sense in certain environments and contexts, in which hand washing should be detected. Similarly to external sensors, this limitation of beacons holds true for all kinds of external cues.
Another approach is to utilize capacitive sensing and to make use of the fact that the metallic piping of water outlets is usually grounded. By measuring the changes in capacitive resistance, events can be detected when the user has touched the outlet or is in contact with the water stream coming out of it \cite{wolling_wettouch_2022}.
However, the assumption that the piping system of wash basins is grounded might not always hold, as these pipes are increasingly built out of plastic materials.

In general, high performances with F1-scores of over 0.9 can be reached \cite{lattanzi_unstructured_2022, zhang_detecting_2021} both in-lab and out-of-lab. However, differences in sensing modalities, environmental cues, and activities included in the datasets make direct comparisons between the existing datasets unfeasible.

\subsection*{Atmospheric sensors in HAR}
Although they are far from omnipresent in Human Activity Recognition (HAR), previous HAR research does use ambient environment sensors such as atmospheric sensors (temperature, pressure, and humidity sensors). For example, in works by De et al. and Bharti et al. \cite{de_multimodal_2015, bharti_human_2019}, atmospheric sensors are utilized to enhance the recognition of complex activities of daily living. Similarly, the sensors are included in a work by Vepakomma et al. \cite{vepakomma_-wristocracy_2015}. These works have in common that they aim to recognize multiple activities of daily living in a living environment ("at home"). By including the ambient environment sensors and additional Bluetooth beacons in their data, they add context to the otherwise harder-to-classify IMU sensor values. They also argue that this sensor type can be used to do on-body localization of wearable devices. For example, the ambient pressure measurement on ankle-worn devices is usually lower compared to wrist-worn devices.

In another research study by Barua et al. \cite{barna_study_2019}, the usefulness of temperature and humidity data is highlighted for the specific activity of using a bathroom, due to the additional context they provide. They find that bathrooms usually deliver higher humidity readings.

Compared to the aforementioned related work, our work similarly uses atmospheric sensors to add context to the IMU recording. However, we specifically show that the humidity and temperature do not only add context about the current location of the user or device but can also be actively employed to directly measure the specific activity of hand washing.

\section{Recording and Evaluation of example hand washing data}
To evaluate the addition of atmospheric sensors, we developed a prototype. We then recorded data using the prototype in a hand-washing detection scenario and analyzed the collected data using visual, statistical, and machine-learning methods.

\subsection{Recording device setup}
We used for our custom wrist-worn wearable prototype a Puck.js embedded units and attached additional atmospheric sensors. We then re-programmed this unit so that it can be used to digitally acquire the integrated IMU's sensor values and the additional atmospheric sensors' values. Since the Puck.js has a limited storage space, we immediately streamed all data to a nearby recording device via Bluetooth low energy (BLE). 

\begin{figure}
    \centering
    \includegraphics[width=1\linewidth]{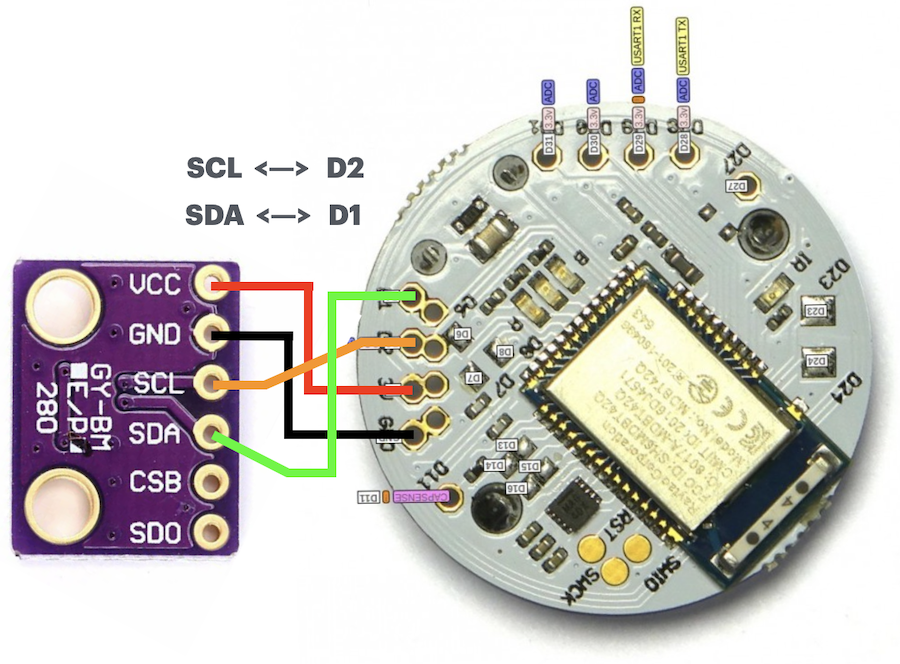}
    \caption{Overview of the wiring needed to attach the BME280 sensor board. The Puck.js provides power to the sensor which can be read via I²C using the D1 and D2 pins. }
    \label{fig:puckBME}
\end{figure}

\begin{figure}
    \centering
    \includegraphics[width=1\linewidth]{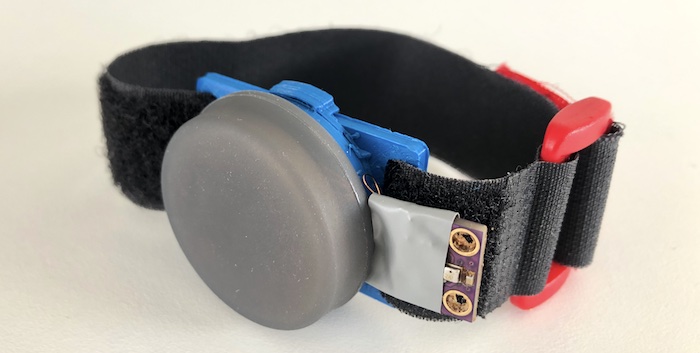}
    \caption{Our prototype consists of a Puck.js with the attached BME280 sensor. A custom 3D-printed enclosure can be mounted on the wrist using the wristband.}
    \label{fig:puckProto}
\end{figure}

Figure \ref{fig:puckProto} shows our initial design. The Puck.js device is powered by a 3V cell battery. The battery life of a single 3V cell battery in our prototype is at least 6 hours, as found per our experiments. In the future, a battery with a larger capacity could be attached as the power source to prolong the battery life and hence increase the maximum recording duration.

\subsection{Example data recording}
To evaluate the usefulness of the addition of the different sensors, we recorded sample data from unscripted hand washes and background activities. We then visually inspected the recordings and trained machine learning classifiers with the collected and annotated data. Finally, we compared the models' performances on different sensor subsets to obtain the influence of each sensor's addition.

We recorded a total of n=10 participants (8m, 2f) for 1 hour per participant. The participants were recorded over multiple weeks, in which highly different external outside conditions were met. The recorded data include recording during sunny and warm days as well as rainy, moist, and cold days. During the 1 hour recording period, each participant washed their hands for a set amount of 4 times. There were no instructions given about the order or duration of hand washing steps, in order not to influence the participants. To also include background data in the recording period, we also included some other activities. During the 1 hour period, the participants were mostly working at their desks. Some recordings also include activities of daily living like cooking pasta and folding laundry. Additionally, we asked each participant to take a walk around the building they were in, which was either an office building or the building they lived in. The walk was constrained to include descending and ascending at least two flights of stairs. Other than this, no further constraints or activities were enforced, so that the data contain mostly realistic daily activities along with the four hand-washing instances. 

To be able to annotate the recordings as accurately as possible, we placed a single Puck.js near designated hand-washing sinks to act as a dedicated BLE beacon. The signals from these beacons were recorded alongside all the sensor data by the wearable units. The proximity to the sink is expected to be correlated to the respective Bluetooth advertisement's received signal strength indicator (RSSI). Added to that, the participants were asked to press the button on the Puck.js device once before starting the hand wash and then once again after finishing the hand wash. This additional data from button presses and beacons is only used for the gathering of ground truth. We restricted the use of beacons to the labeling because to gather this data, beacons would have to be placed at every sink, the user would wash their hands at. Similarly, in the real application, button presses before and after each hand-wash would nullify the need for automated hand-washing detection, hence the button presses are also only used for ground truth annotation.

We recorded the sensor data from the integrated and externally attached sensors as shown in Table \ref{tab:sensors}.

\begin{table}[]
\begin{tabular}{|l|l|l|l|}
\hline
sensor type & axes & device  & sampling frequency \\ \hline
Accelerometer    & 3    & Puck.js & 52 Hz          \\
Gyroscope         & 3    & Puck.js & 52 Hz          \\
Humidity          & 1    & BME280  & 1 Hz           \\
Temperature       & 1    & BME280  & 1 Hz           \\
Atm. Pressure     & 1    & BME280  & 1 Hz           \\ \hline
\end{tabular}

\caption{The variety of sensors used for recording data in our experiment, which are included in the Bosch BME280 sensor board and the Puck.js sensor board. The latter collects all sensor readings to be forwarded via Bluetooth Low Energy.\label{tab:sensors}}

\end{table}

The recorded data was then manually annotated in Label Studio \cite{tkachenko_label_2020}.
For the annotation, the button presses and the proximity values to the Bluetooth beacons were visualized alongside the IMU sensor values and the values of the atmospheric sensors. This enabled us to accurately label the hand-washes.

In addition to recording sensor data, we also noted down the recording day's weather conditions for each recording, i.e. mean temperature, mean relative humidity, condition (rainy, sunny, cloudy), and mean air pressure. We include this meta-information in the dataset. Overall, we recorded on 6 different days, with rainy to sunny conditions, temperatures from 13 to 21°C, humidity between 63.5 and 89, and pressure in the range of 996.7 to 1007.7\,hPa. However, since the data was recorded indoors, the outside weather conditions likely did not impact the results severely. Also, we found that the readings of the recorded atmospheric sensors are not well-calibrated, i.e. the absolute values are not always correct. In contrast, the relative values are reliable, showing similar changes to external sensors to which we compared them.

\subsection{Evaluation of collected data}
\begin{figure}
    \centering
    \includegraphics[width=0.95\linewidth]{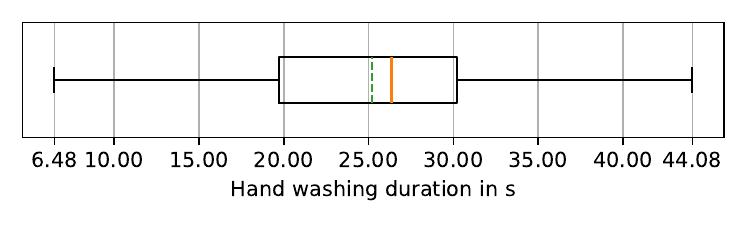}
    \caption{Statistics for the duration in seconds of all 43 recorded hand washes. The box plot shows the median (red solid line), mean (green dashed line), quartiles (box extents), and minimum and maximum (whiskers).}
    \label{fig:hw_stats}
\end{figure}

In total, we collected 43 instances of hand washing. In Fig. \ref{fig:hw_stats} we show the distribution of the hand-washing duration for all hand-washes contained in the recorded data. The mean (25.19s), median (26.34s), and the quartiles of the duration of a hand-washing instance are between 20s and 30s. However, the minimum (6.48s) and the maximum (44.08s) deviate from the mean. In total, the recorded data has a length of 10 hours and 3.5 minutes.
The contained hand washing has a total duration of 17.85 minutes.
Thus, the collected dataset is highly imbalanced with the null class making up 97\,\% of the data.
However, we argue that this is a desirable feature of our dataset, as hand washing is also scarce in the real world, where it only makes up a tiny percentage of a potential user's daily activities.

\subsection{Visual representation}
To inspect the additional sensors' usefulness for our tasks, we created visual representations, which will be shown and discussed in this chapter.

\begin{figure*}
  \centering
  \includegraphics[width=\textwidth]{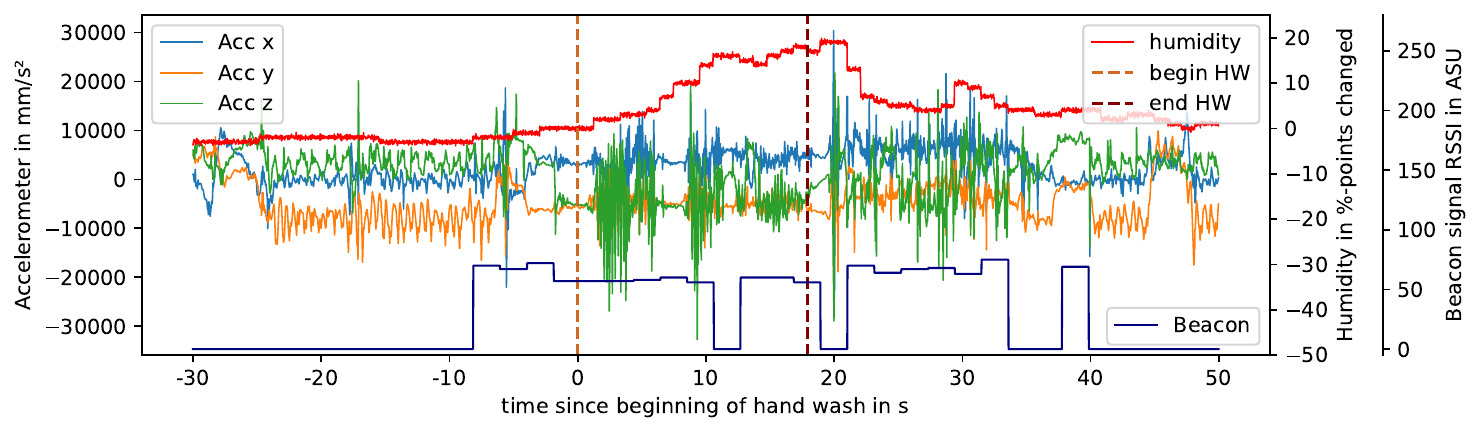}
  \caption{Example accelerometer and humidity sensor plot of one hand washing (HW in legend) instance. We can observe the humidity rising once the participant starts washing their hands at time t = 0. Additionally, the signal RSSI received from a Bluetooth beacon placed at the sink is shown (navy blue). The Beacon signal was only used for labeling and is non-zero while the participant remained near the sink, e.g. to dry off the hands after washing.}
  \label{fig:exampleHW}
\end{figure*}

One example of hand washing is displayed in Fig. \ref{fig:exampleHW}, which shows the accelerometer values and the humidity change compared to the start of hand washing. In the plot, we can observe the participant walking to the sink, then starting to wash their hands, and then drying off their hands before finally walking again. The humidity level rises shortly after the hand washing is started by the participant, and begins to fall one the tap is turned off and the participant is drying off the hands.

\begin{figure*}
\centering
\subfloat[Humidity: After the end of the hand washing, the humidity decreases. The decrease is not as sudden as the prior increase, which occurs around the start of the handwashing.]{\includegraphics[trim={7mm 0 0 0},clip,width=0.32\linewidth]{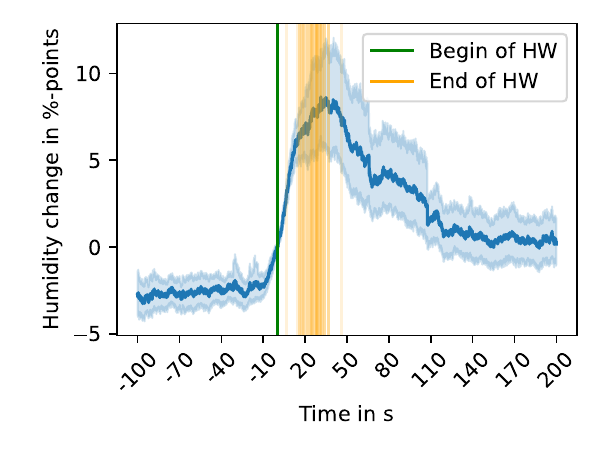}} \hfill
\subfloat[Temperature: The readings tend to decrease slightly (note the Y axis scale) when entering the bathroom or kitchen, and rises after leaving it again. ]{\includegraphics[trim={7mm 0 0 0},clip,width=0.32\linewidth]{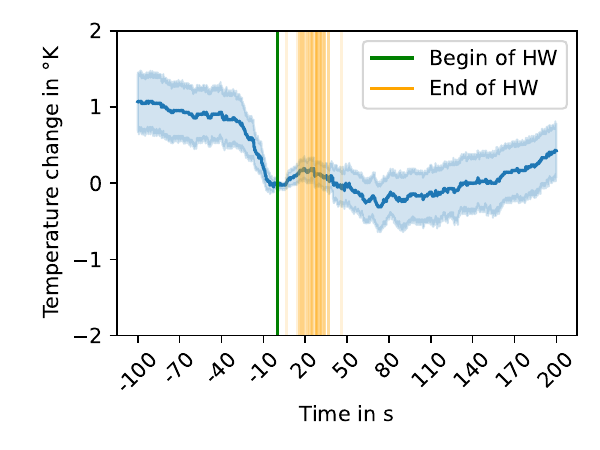}}\hfill
\subfloat[Air Pressure: The pressure sensor's readings stay relatively constant during the hand washing. The larger deviation after the hand washing comes from one participant who took a walk up the stairs of the building shortly after washing.]{\includegraphics[trim={7mm 0 0 0},clip,width=0.32\linewidth]{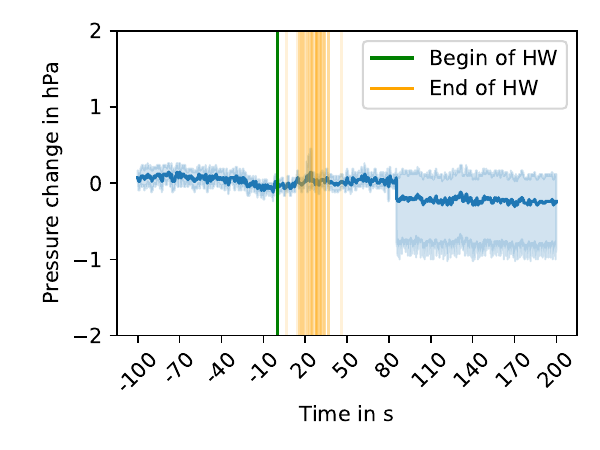}}
\caption{Response of the (a) humidity, (b) temperature, and (c) pressure sensors to hand washing, averaged (in dark blue) over all recorded hand washes, with bootstrapped 95\% confidence interval (in light blue). The start of the hand washing is marked with a green vertical line. The yellow vertical lines mark the respective ends of all the handwashing instances.}
\label{fig:hw_resp}
\end{figure*}

The humidity change after starting to wash hands is shown in Fig. \ref{fig:hw_resp} (a). As expected, the measured humidity begins to increase around the moment at which the participant starts their handwashing procedure. The signal then peaks some seconds after the start of the hand wash, depending on the duration of the hand wash. It can also be observed that after a hand wash, the sensor values only slowly decrease again until they reach the same level that they were at before starting the hand wash. During the hand wash, the humidity usually monotonously rises until the hand wash is finished and the participant starts to dry off their hands. Interestingly, the humidity already starts to rise before the hand wash is started, which is probably due to the room in which the sink is located having a slightly higher humidity in comparison to other rooms, which would align well with the findings in \cite{barna_study_2019}.

The temperature response of washing one's hands is displayed in Fig. \ref{fig:hw_resp} (b). It seems that the hand-washing has no immediate effect on the measured temperature, as the temperature remains stable during the wash. However, before the hand wash, the temperature decreases and afterward it increases again. We assume that this is due to the bathroom/kitchen being colder than the other rooms the participants were in before hand washing. Therefore, although there seems to be no change during the hand washing, the temperature sensor might deliver context about the room in which a user might be staying.

As expected, the pressure sensor's reading stays almost perfectly constant before, during, and after washing one's hands. Therefore, we did not include a visual representation of it in this report. It follows that the pressure sensor readings are unlikely to have a large effect on hand washing detection accuracy. However, when the pressure sensor shows a high rate of change, we can also rule out hand washing as the user's current activity. In that sense, there is still an expected benefit of including it in the data, e.g. filtering out the ``stairs'' activity.

\subsection{Machine Learning Experiments}
After the visual inspection of the recorded data showed promising correlations between the newly added humidity sensor and the hand washing labels, we trained random forest classifiers on the data. The goal of the machine learning experiment was not to maximize the performance but rather to explore the importance of the multi-modal sensor setup. Specifically, each sensor's impact on the prediction performance of the trained classifier was to be evaluated.

Hence, we trained models using different sensor configurations in a small-scale ablation study and recorded the performance of the random forest classifier for each sensor set. The sensor subsets we used are listed in Table \ref{tab:sensors}. The different sensor combinations are used to measure the impact of each sensor's addition or removal from the dataset. 

\begin{table}[]
\begin{tabular}{l|l}
sensors                   & abbreviation \\ \hline
Accelerometer             & A            \\
Acc + H. + T. + P             & A+HTP       \\
Accelerometer + Gyroscope & AG           \\
Acc. + Gyro. + Humidity   & AG+H         \\
Acc. + Gyro + Temperature & AG+T         \\
Acc + Gyro + Pressure     & AG+P         \\
Acc + Gyro + H. + T. + P. & ALL         
\end{tabular}
\caption{Sensor subsets used for training and evaluation. H stands for humidity,  T for temperature and P for pressure.}
\label{tab:sensors}
\end{table}

We trained and evaluated the models on non-overlapping 2.5s and 5s-long sliding windows using the feature set in the enumeration below:
\begin{itemize}
    \item mean
    \item standard deviation
    \item minimum
    \item maximum
    \item slope (last value - first value)
    \item median
    \item inter-quartile-range
    \item first quartile
    \item third quartile
    \item average crossings (times the signal crosses the mean)
    \item skewness
    \item kurtosis
\end{itemize}
We chose this window length in line with other hand-washing detection literature (\cite{burchard_washspot_2022}: 3s, \cite{cao_leveraging_2023}: 6s). The feature set was designed to represent the findings of the previously mentioned visual inspection of the data. We included the usual statistical features of mean, std, min, max, and interquartile ranges for each sensor axis. Additionally, we added the slope of all sensors. One could also go a step further and add features in the frequency domain, but for the basic evaluation of the recorded data, we did not include this type of feature in this work.

To receive an estimate of the participant-independent performance, we explored a leave-one-participant-out (LOPO/LOSO) cross-validation. Additionally, we used per-participant personalized models, to show the effect on single participants. For the personalized model, we trained and evaluated on a train-test split, using 33\,\% of the user's data for testing and the rest for training the model.

We used a seeded random number generator and repeated the experiment five times to reduce the randomness of the outcome.

\section{Results}
\label{sec:results}
The resulting F1-scores of our machine learning experiments are displayed in Table \ref{tab:results} for both the LOSO-split and for the personalized machine learning.

We also included the maximal baseline performance of a dummy classifier to better relate the achieved F1 scores. The dummy classifier performance is chosen as the best-performing dummy classifier performance in any of the splits for the relevant task and window size. Hence we over-estimate the dummy performance and find an upper bound for the chance level.

The trained model's performance is significantly above the chance level, with an average F1 score of 69.15\,\% for the participant-independent task with 5\,s windows, and an average F1 score of 85\,\% for the personalized task. This performance shows that the classifiers were able to detect hand washing from the recorded background activities with high precision and recall. On the other hand, the higher performance on the personalized task also shows that hand washing is highly user-dependent. This could be explained by user-specific rituals and patterns of hand-washing.

However, the results do not yield the expected result, that the addition of the atmospheric sensors boosts machine learning performance. Unlike the visual inspection proposed, the selected sensor subset did not have a measurable impact on the performance. For the participant-independent task, the "classic" combination of accelerometer and gyroscope reached the highest performance. In the personalized task, the atmospheric sensors in combination with the accelerometer yielded the highest performance, with a tiny lead.

\begin{table*}[]
\begin{tabular}{c|cllllllll}
\multicolumn{1}{l|}{sensors}  & window size & A     & A+HTP & AG    & AG+H  & AG+T  & AG+P  & ALL   & Chance \\ \hline
\multirow{2}{*}{LOSO}         & 2.5s        & 0.663 & 0.636 & \textbf{0.679} & 0.675 & 0.673 & 0.676 & 0.667 & 0.111  \\
                              & 5s          & 0.681 & 0.683 & \textbf{0.704} & 0.693 & 0.691 & 0.702 & 0.688 & 0.077  \\ \hline
\multirow{2}{*}{Personalized} & 2.5s        & 0.819 & 0.846 & 0.844 & 0.830 & 0.848 & \textbf{0.857} & 0.849 & 0.188  \\
                              & 5s          & 0.854 & \textbf{0.860} & 0.859 & 0.848 & 0.844 & 0.833 & 0.852 & 0.076 
\end{tabular}
\caption{Resulting \textbf{F1 scores} of the machine learning experiments. No sensor set performed significantly better than the others. In the participant-independent task, the Accelerometer \& Gyroscope performed best, while in the personalized task, the Accelerometer \& Gyroscope together with Pressure, and Accelerometer \& all atmospheric sensors took the lead. Note that these results also show that the dataset is more challenging than previous work, which sometimes reaches F1 scores over 0.9 (see Section \ref{sec:rel_work}). }
\label{tab:results}
\end{table*}

\section{Discussion}
The results show that the addition of the atmospheric sensors can provide additional context. In the visual inspection (see Fig. \ref{fig:exampleHW} and Fig. \ref{fig:hw_resp} (a)) of the collected data, it becomes apparent that the increase of humidity near the running tap can be measured reliably by the used sensors. 
This could be helpful in visual inspection and retrospective annotation of wearable sensor data where hand washing might occur.  
Thus, we would have expected, that the hand washing detection performance is best with all sensors or at least the humidity sensor included. There could be several reasons why this is not the case in our experiment, with the manually crafted features not taking advantage of this additional sensor's potential as the most likely explanation. This points to the need for more specific features for the slower and relative changes in humidity readings in particular.

We also visualized the effect of hand washing on the temperature recorded by the sensors. The change of measured temperature in proximity to the running tap is found to be negligible, whilst the change of room is consistently picked up in the sensor data. 

While we can think of other applications for the atmospheric pressure sensor, we do not see a large benefit in using it for hand washing detection. While the pressure sensor probably helps to filter out activities involving altitude changes, it likely does not provide additional discriminatory performance against hand washing for most activities.

In future work, more data and more background activities could be recorded using the same system, which can be replicated as it is open-source. A different machine learning paradigm, like end-to-end deep learning, could also be employed to more precisely capture the characteristic patterns of hand washing for the additional sensors.

Our developed prototype itself is easy to build and records the data reliably.
One limitation of the prototype is its reliance on a nearby recording device. For our experiments, we used a laptop, that was carried around by a researcher following the respective participant everywhere. In the future, if we keep the prototype based on the Puck.js, a smartphone application can be used in a similar fashion. A smartphone with the application could be carried around by participants in their pockets, without creating additional disturbance.

Another limitation lies in the nature of the humidity sensor itself. Due to it measuring the humidity in the air, the sensor values only rise slowly, once the user starts washing their hands. They also decrease over a long period of time, inducing some lag that current features do not account for perfectly. The added sensor thus provides a limited benefit if we want to detect the exact onset or offset of the hand-washing activity, due to its slow adaptation. For offline analysis, time-shifting the humidity signal by an appropriate amount of milliseconds could be beneficial. For online analysis, no such workaround seems obvious. Therefore, the humidity sensor is only an addition to the much more frequently used IMU sensor and likely should not be seen as a replacement for the latter.

\section{Conclusions}
This work showcases an open-source implementation of a hand-washing detection system that uses sensing capabilities that go beyond the usual sensing modalities that are used in wearable-based Human Activity Recognition (HAR). The developed prototype and its software are evaluated in a feasibility study, recording the hand-washing behavior of 10 participants. By analyzing the recorded data visually, it is shown that the addition of some of the modalities, in particular humidity, can provide distinctive readings for improving the hand washing detection performance. Concrete evidence of a large effect on machine learning performance must be provided by further research on more data, and more specific features for these new modalities. Such features should e.g. better capture the observed pattern of rising humidity after most hand washing onsets. 

Applications for hand washing detection, hand washing duration predictions, hand washing quality prediction, and similar problems are manifold and can be found in the domains of industry, healthcare, mental health, and more. 

Based on our work, we argue that especially atmospheric sensors should be considered for wearable devices aimed at reliably detecting hand washing and distinguishing it from other activities. In general, while atmospheric sensors have been proposed to be used for HAR in the past, recent research has usually neglected them. We argue that there are still more potential applications of them to be explored in the future. The open question of extracting distinctive features that describe the relative humidity rise and drop before and after hand washing is the topic of such work in progress.

The firmware files for the Puck.js, the data recording script, the recorded data, as well as the code used to produce our results, can be found in our GitHub repository\footnote{\url{https://github.com/kristofvl/wearPuck}}.

\begin{acks}
This project is funded by the Deutsche Forschungsgemeinschaft (DFG, German Research Foundation) – 425868829 and is part of Priority Program SPP2199 Scalable Interaction Paradigms for Pervasive Computing Environments.
\end{acks}

\bibliographystyle{ACM-Reference-Format}
\balance
\bibliography{main}

\end{document}